\documentclass[preprint]{revtex4}
\usepackage{graphicx}
\begin{document}
\draft
%@twocolumnfalse\endcsname

\title[]{Kinetically-controlled thin-film growth of layered
$\beta$- and $\gamma-$Na$_{x}$CoO$_{2}$ cobaltate}

\author{J. Y. Son, Bog G. Kim and J. H. Cho}

\affiliation{ RCDAMP and Department of Physics, Pusan National University, Busan 609-735, Korea}

\begin{abstract}

We report growth characteristics of epitaxial
$\beta$-Na$_{0.6}$CoO$_{2}$ and $\gamma$-Na$_{0.7}$CoO$_{2}$ thin
films on (001) sapphire substrates grown by pulsed-laser
deposition. Reduction of deposition rate could change
structure of Na$_{x}$CoO$_{2}$ thin film from $\beta$-phase
with island growth mode to $\gamma$-phase with layer-by-layer
growth mode. The $\gamma$-Na$_{0.7}$CoO$_{2}$ thin film exhibits
spiral surface growth with multiterraced islands and highly
crystallized texture compared to that of the
$\beta$-Na$_{0.6}$CoO$_{2}$ thin film. This heterogeneous
epitaxial film growth can give opportunity of strain effect of
physical properties and growth dynamics of Na$_{x}$CoO$_{2}$ as
well as subtle nature of structural change.


\end{abstract}

\maketitle

Families of sodium cobalt oxide Na$_{x}$CoO$_{2}$ are attractive
materials due to their large thermoelectric power and low
resistivity, which can be applied for thermoelectric
applications\cite{son1,son2}. The large thermoelectric power is
attributed to the spin entropy from the low-spin state of Co
ion\cite{son3}. Occurrence of superconductivity of
Na$_{0.35}$CoO$_{2}$$\cdot$1.3H$_{2}$ with T$_{C}$ about 5 K and
rich phase diagrams of Na$_{x}$CoO$_{2}$ with respect to x also
inspire many theoretical and experimental interests of novel
ground states due to its two-dimensional transition-metal oxide
triangular lattice\cite{son4}.

Until now, physical properties of
single crystal and powder of Na$_{x}$CoO$_{2}$ have been widely
studied\cite{son5,son6,son7,son8,son9} but there have been no
rigorous reports about thin film studies of Na$_{x}$CoO$_{2}$.
Since the thin film is interesting in its own due to strain effect
as well as the fact that most application can be done in thin film
structure, the thin film growth of Na$_{x}$CoO$_{2}$ can give an
opportunity of manipulation of physical properties by changing
strain field using different substrates and growth parameters
(substrate temperature, oxygen partial pressure, etc). In this
study, we report the thin film growth of differently structured
epitaxial Na$_{x}$CoO$_{2}$ by Pulsed Laser Deposition (PLD) on
(001) sapphire.

The crystal structure of Na$_{x}$CoO$_{2}$
consists of two-dimensional triangular CoO$_{2}$ layers of
edge-sharing CoO$_{6}$ octahedra separated by an insulating layer
of Na$^{+}$ ions. There are four known phases of Na$_{x}$CoO$_{2}$
with slightly different structures such as $\alpha$-,
$\alpha$$\prime$-, $\beta$-, and $\gamma$-  phases of
Na$_{x}$CoO$_{2}$ distinguished by stacking order of CoO$_{2}$
layers and Na-O environments. Figure 1 shows schematic structures
of $\beta$-Na$_{x}$CoO$_{2}$ and $\gamma$-Na$_{x}$CoO$_{2}$ in a-c
plane. The $\beta$-phase has a monoclinic unit cell with a space
group symmetry of C2/m and lattice constants of a = 4.902, b =
2.828, c = 5.720 $\AA$ and $\beta$ = 105.96$^{\circ}$. The
$\gamma$-phase has a hexagonal structure with a space group
symmetry of P6$_{3}$/mmc and lattice constants of a = 2.840 and c =
10.811 $\AA$\cite{son10,son11,son12}. In
$\gamma$-Na$_{x}$CoO$_{2}$, in-plane direction of CoO$_{6}$
octahedron in CoO$_{2}$ layer is alternating with the nearest
CoO$_{2}$ layers, whereas, in $\beta$-Na$_{x}$CoO$_{2}$, the
direction is parallel with the nearest CoO$_{2}$ layers. These
structural varieties of Na$_{x}$CoO$_{2}$ imply that the bulk
modulus along the plane is small and the stress is an important
parameter for the growth of the epitaxial thin films because of
various possible stacking structures of CoO$_{2}$ layer and
possibility of stacking fault as seen in bulk Na$_{x}$CoO$_{2}$.

Na$_{x}$CoO$_{2}$ thin films have been grown by PLD method. The
Na$_{0.8}$CoO$_{2}$ target, which was used in PLD, was prepared by a
conventional solid-state reaction method. Na$_{2}$CO$_{3}$ (99.995 $\%$) and
Co$_{3}$O$_{4}$ (99.998 $\%$) were mixed in molar ratio of Na : Co = 0.8 :
1.0 and the mixed powder was pressed into pellet and calcined at
750$^{\circ}$C for 12 h. The calcined pellet was reground, pressed
into pellet, and sintered at 850$^{\circ}$C for 24 h. (001)
sapphire substrate was used for the thin film growth and the
substrate temperature of optimal thin film growth was
480$^{\circ}$C. A frequency tripled (355 nm, $\sim$ 2 J/cm$^{2}$)
Nd:YAG laser was used for the deposition and the distance between
target and substrate was $\sim$ 4 cm. The deposition rates in the
range of 0.02 $\sim$ 0.2 $\AA$ /pulse were controlled by
repetition of laser pulse and an eclipse method, in which a shadow
mask was placed between the target and the substrate. The energy
of an adatom was reduced by the eclipse method because direct
high-energy particles would be rejected by the shadow mask.
Optimal oxygen pressure of 400 mTorr was maintained during the
deposition.

The $\gamma$-Na$_{0.7}$CoO$_{2}$ thin film was grown
with layer-by-layer growth mode by the low deposition rate of 0.02
$\AA$ /pulse using the eclipse method on (001) sapphire
substrates. When the deposition rate was increased, the mixed
phase of $\beta$- and $\gamma$-Na$_{x}$CoO$_{2}$ was observed. By
the condition of the high deposition rate of 0.2 $\AA$ /pulse,
only the $\beta$-Na$_{0.6}$CoO$_{2}$ thin film was grown with island
growth mode. For the structural determinations of $\beta$- and
$\gamma$-Na$_{x}$CoO$_{2}$ thin films, x-ray diffraction data were
obtained by conventional laboratory x-ray as well as synchrotron
x-ray sources at 5C2 in Pohang Light Source. The thicknesses of
$\beta$- and $\gamma$-Na$_{x}$CoO$_{2}$ thin films were $\sim$1000
and $\sim$2000 $\AA$, respectively, determined from the
cross-sectional images of scanning electron microscope (SEM). The
tentative compositions of $\beta$-Na$_{0.6}$CoO$_{2}$ and
$\gamma$-Na$_{0.7}$CoO$_{2}$ were obtained by energy dispersive
x-ray spectrometer (EDS). The surface morphologies and
topographies of $\beta$-Na$_{0.6}$CoO$_{2}$ and
$\gamma$-Na$_{0.7}$CoO$_{2}$ thin films were observed by SEM and
atomic force microscope (AFM).

Figure 2 (a) shows the x-ray
diffraction pattern of the epitaxially grown (00$\it{l}$)
$\beta$-Na$_{0.6}$CoO$_{2}$ thin film on (001) sapphire substrate.
The full width at half maximum (FWHM) of the (001) rocking curve
is about 1.4$^{\circ}$. The out-of-plane lattice constant of 5.459
$\AA$ was obtained from the (00$\it{l}$) peaks and this value is
slightly smaller than that of bulk $\beta$-Na$_{x}$CoO$_{2}$. To
check the in-plane orientation, we performed x-ray scattering with
$\Phi$-scan geometry. Figure 2 (b) shows $\Phi$-scan of
($\bar{1}$12) peak of $\beta$-Na$_{0.6}$CoO$_{2}$ thin film and
(104) peak of the sapphire substrate. The ($\bar{1}$12) peaks of
$\beta$-Na$_{0.6}$CoO$_{2}$ thin film are located 30$^{\circ}$ off
from the (104) plane of sapphire substrate. The sixfold symmetry
of ($\bar{1}$12) peaks from the twinned grains is observed,
representing that the a-lattice of $\beta$-Na$_{0.6}$CoO$_{2}$
thin film is oriented along the hexagonal symmetry of a-axis
lattice of (001) sapphire substrate. The FWHM of in-plane
($\bar{1}$12) peak is equal to 1.3$^{\circ}$. From the 2$\theta$ values
of the (111) and the ($\bar{1}$12) peaks, we obtain a monoclinic
a-axis lattice constant of 4.886 $\AA$ and b-axis lattice constant
of 2.818 $\AA$ that are close to the bulk lattice constant (a =
4.902, b = 2.828 $\AA$) of $\beta$-Na$_{0.6}$CoO$_{2}$. To check a
strain of the film, $\it{h}$, $\it{k}$, and $\it{l}$ scans near
the ($\bar{1}$12) plane were performed. The data indicated that
the strain along the in-plane direction was low estimated from the
small FWHM of $\it{h}$, $\it{k}$ scans and the strain along the
out-of-plane direction was high resulting from the large FWFM of
$\it{l}$ scan. To investigate the growth rate effects, we have
grown Na$_{x}$CoO$_{2}$ thin film on (001) sapphire substrates
using the same target with the low deposition rate of 0.02 $\AA$
/pulse by the eclipse method.

Figure 3 (a) shows the x-ray
diffraction pattern of the epitaxially grown (00$\it{l}$)
$\gamma$-Na$_{0.7}$CoO$_{2}$ thin film on (001) sapphire
substrate. The FWHM of the (002) rocking curve is about
0.8$^{\circ}$ and this FWHM of $\gamma$-Na$_{0.7}$CoO$_{2}$ thin
film is smaller than that of $\beta$-Na$_{0.6}$CoO$_{2}$ thin
film, indicating that the (00$\it{l}$)
$\gamma$-Na$_{0.7}$CoO$_{2}$ thin film shows better crystallized
texture along the out-of-plane direction. The c-lattice constant
of 10.848 $\AA$ was obtained from the (002) peak. Figure 3 (b)
shows $\Phi$-scan of (104) peak of the epitaxial
$\gamma$-Na$_{0.7}$CoO$_{2}$ thin films and (104) peak of the
(001) sapphire substrate. The sixfold symmetry represents the
hexagonal structure of $\gamma$-Na$_{0.7}$CoO$_{2}$ thin film with
twins. The FWHM of in-plane (104) peak is equal to
0.6$^{\circ}$ and this indicates that the
$\gamma$-Na$_{0.7}$CoO$_{2}$ thin film also shows better
crystallized texture along the in-plane direction than that of the
$\beta$-Na$_{0.6}$CoO$_{2}$ thin film. From the 2$\theta$ value of (104)
peak, we obtain a hexagonal a-axis lattice constant of 2.812 $\AA$
that is similar to the bulk $\gamma$-Na$_{0.7}$CoO$_{2}$. 
Thus, $\gamma$-Na$_{0.7}$CoO$_{2}$ film grown by the low
deposition condition shows better-crystallized texture both
in-plane and out-of-plane direction.

Figure 4 (a) and (b) show the
AFM images of epitaxial $\beta$-Na$_{0.6}$CoO$_{2}$ and
$\gamma$-Na$_{0.7}$CoO$_{2}$ thin films, respectively. The large
grains with typical tetrahedral islands are observed in
$\beta$-Na$_{0.6}$CoO$_{2}$ thin film. The shape indicates that
the diffusion length along the surface is short during deposition
and the kinetics of the step is limited locally resulting from the
high deposition rate. The root mean square (rms) surface roughness
has the large value of 220 $\AA$ (Figure 4 (c) and (d)). However,
the spiral patterns with multi-terraces are observed in
$\gamma$-Na$_{0.7}$CoO$_{2}$ thin film, and this indicates that
epitaxial $\gamma$-Na$_{0.7}$CoO$_{2}$ thin film was grown with
atomically flat surface by monolayer steps. The surface roughness
of $\gamma$-Na$_{0.7}$CoO$_{2}$ thin film is extremely smooth with
the rms roughness of 8 $\AA$. Figure 4 (e) and (f) show the
sectional contour graphs of the $\#$1-line and the $\#$2-line in
$\gamma$-Na$_{0.7}$CoO$_{2}$ thin film. The large terraces with a
width of 1000 $\sim$ 2000 $\AA$ are observed and the terrace
heights are nearly half or same of one lattice unit along the
c-axis. This large width of the terraces represents that the
surface diffusion of adatoms along the surface is long enough and
the kinetics of the step is widespread due to the low deposition
rate inhibiting from frequent nucleation of adatoms\cite{son13}.
Therefore, $\gamma$-Na$_{0.7}$CoO$_{2}$ thin film shows
layer-by-layer growth following the ideal step flow growth mode,
closely.

In conclusion, we have grown the epitaxial
$\beta$-Na$_{0.6}$CoO$_{2}$ and $\gamma$-Na$_{0.7}$CoO$_{2}$ thin
films on the (001) sapphire substrate
deposited by PLD method. On the low deposition rate of 0.02 $\AA$
/pulse, the $\gamma$-Na$_{0.7}$CoO$_{2}$ thin film was grown with
layer-by-layer growth. When a deposition rate was increased, a
mixed phase of $\beta$- and $\gamma$-Na$_{x}$CoO$_{2}$ was
observed. By the condition of the high deposition rate of 0.2
$\AA$ /pulse, the $\beta$-Na$_{0.6}$CoO$_{2}$ thin film was grown
with tetrahedral islands. The $\gamma$-Na$_{0.7}$CoO$_{2}$ thin
film exhibits spiral surface growth with multi-terraces and highly
crystallized texture. These experimental demonstrations and
controllability could provide opportunities of strain effects of
Na$_{x}$CoO$_{2}$ and physical properties of thin films and growth
dynamics of heterogeneous epitaxial thin films.

This work was supported by Korea Research Foundation Grant.
(KRF-2004-005-C00045).

\newpage

\begin{figure}[c!]
\includegraphics[width=15cm]{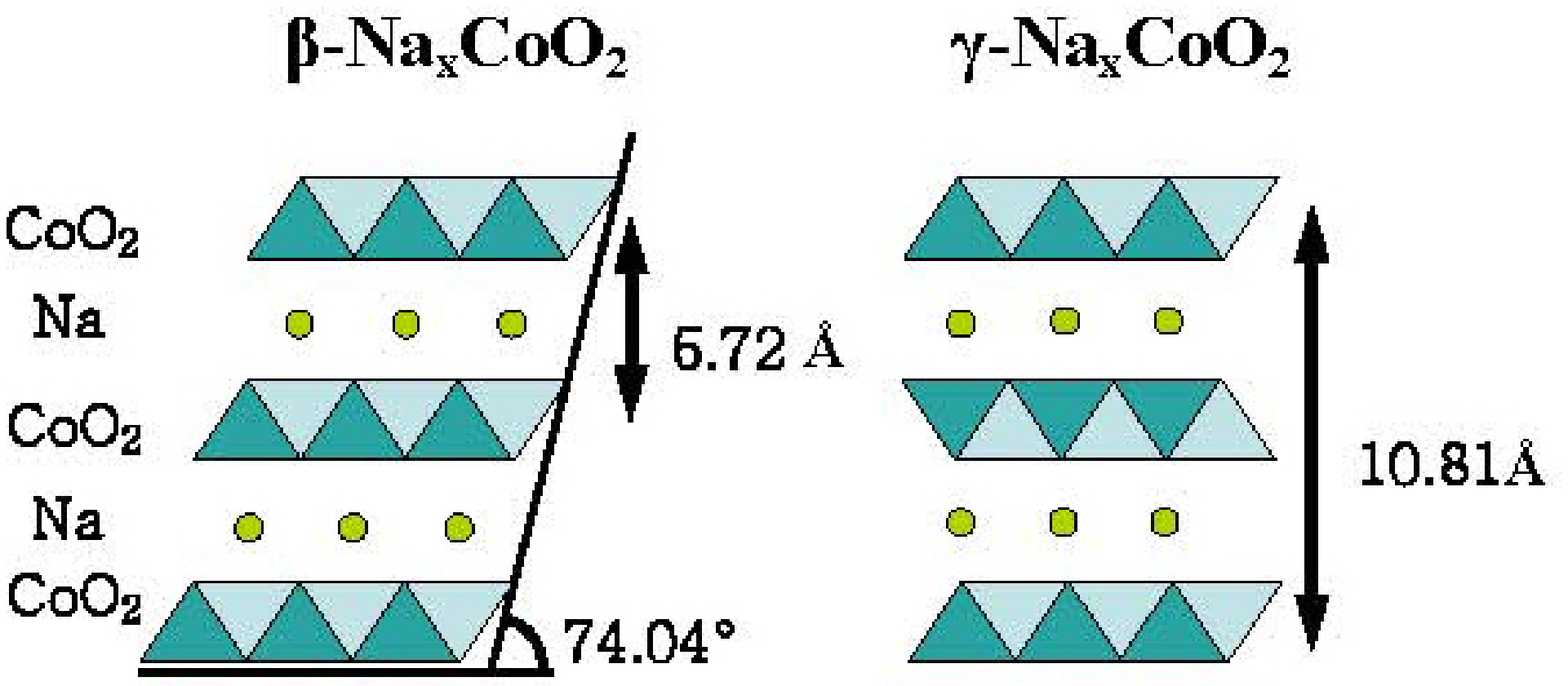}
\caption{ Schematic structures of $\beta$-Na$_{x}$CoO$_{2}$ and
$\gamma$-Na$_{x}$CoO$_{2}$. Rhombus and filled circle symbolizes a
CoO$_{6}$ octahedron and Na ions, respectively.
$\beta$-Na$_{x}$CoO$_{2}$ has a monoclinic structure with $\beta$
= 105.96$^{\circ}$. In $\gamma$-Na$_{x}$CoO$_{2}$, in-plane
direction of CoO$_{6}$ octahedron in CoO$_{2}$ layer has the
opposite direction of in-plane direction of CoO$_{6}$ octahedron
in the nearest CoO$_{2}$ layers. }
%\label{f1}
\end{figure}

\newpage
\begin{figure}[c!]
\includegraphics[width=15cm]{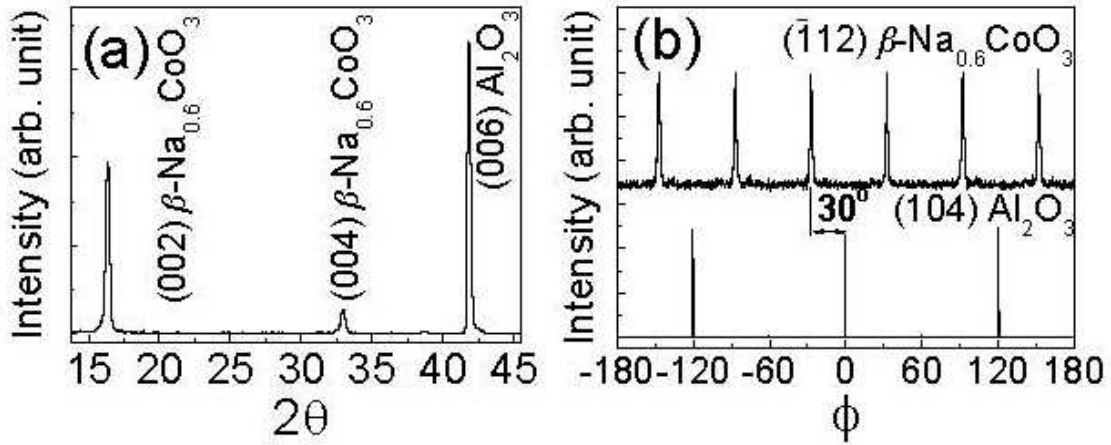}
\caption{(a) X-ray diffraction pattern of the epitaxial
$\beta$-Na$_{0.6}$CoO$_{2}$ thin films. (b) $\Phi$-scan of
($\bar{1}$12) peaks of the epitaxial $\beta$-Na$_{0.6}$CoO$_{2}$
thin films and (104) peaks of (001) sapphire substrate. }
%\label{f1}
\end{figure}

\newpage
\begin{figure}[c!]
\includegraphics[width=15cm]{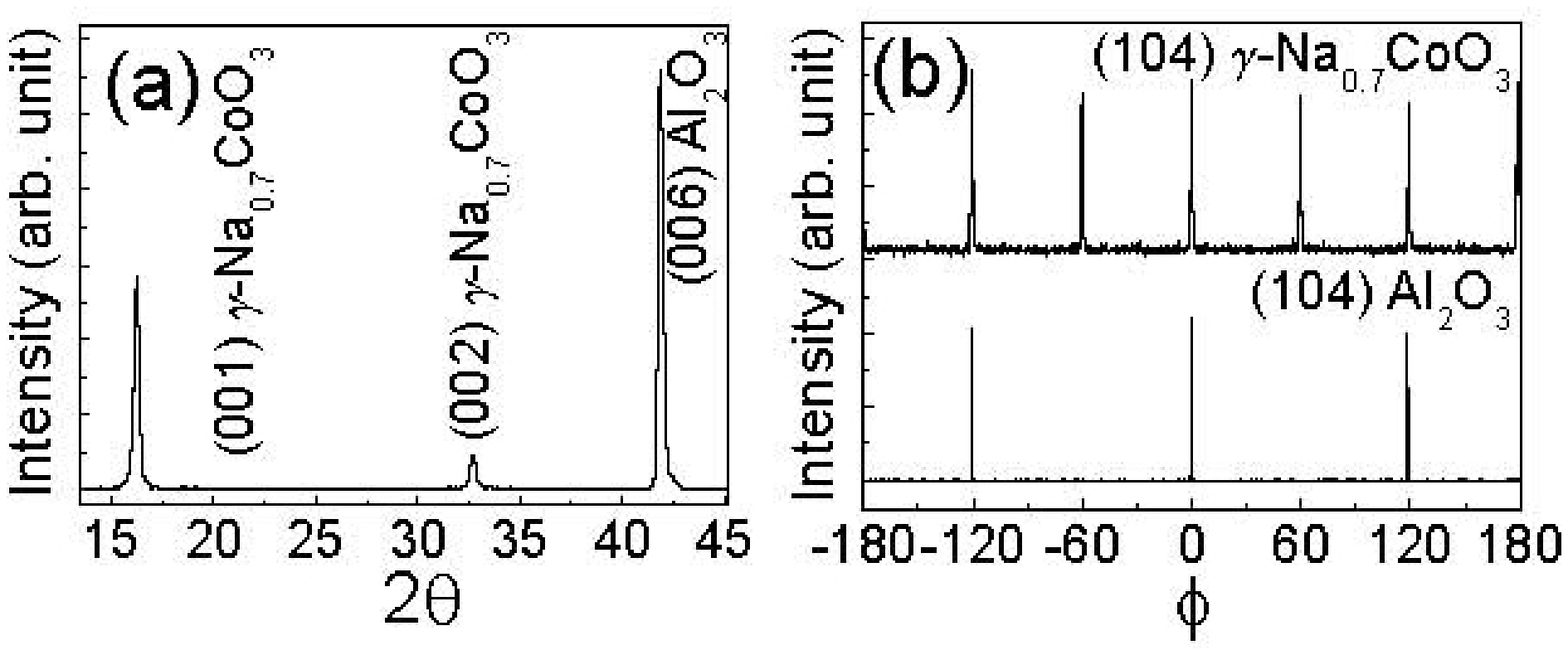}
\caption{(a) X-ray diffraction pattern of the epitaxial
$\gamma$-Na$_{0.7}$CoO$_{2}$ thin films. (b) $\Phi$-scan of (104)
peaks of the epitaxial $\gamma$-Na$_{0.7}$CoO$_{2}$ thin films and
(104) peaks of (001) sapphire substrate.}
%\label{f1}
\end{figure}

\newpage
\begin{figure}[c!]
\includegraphics[width=15cm]{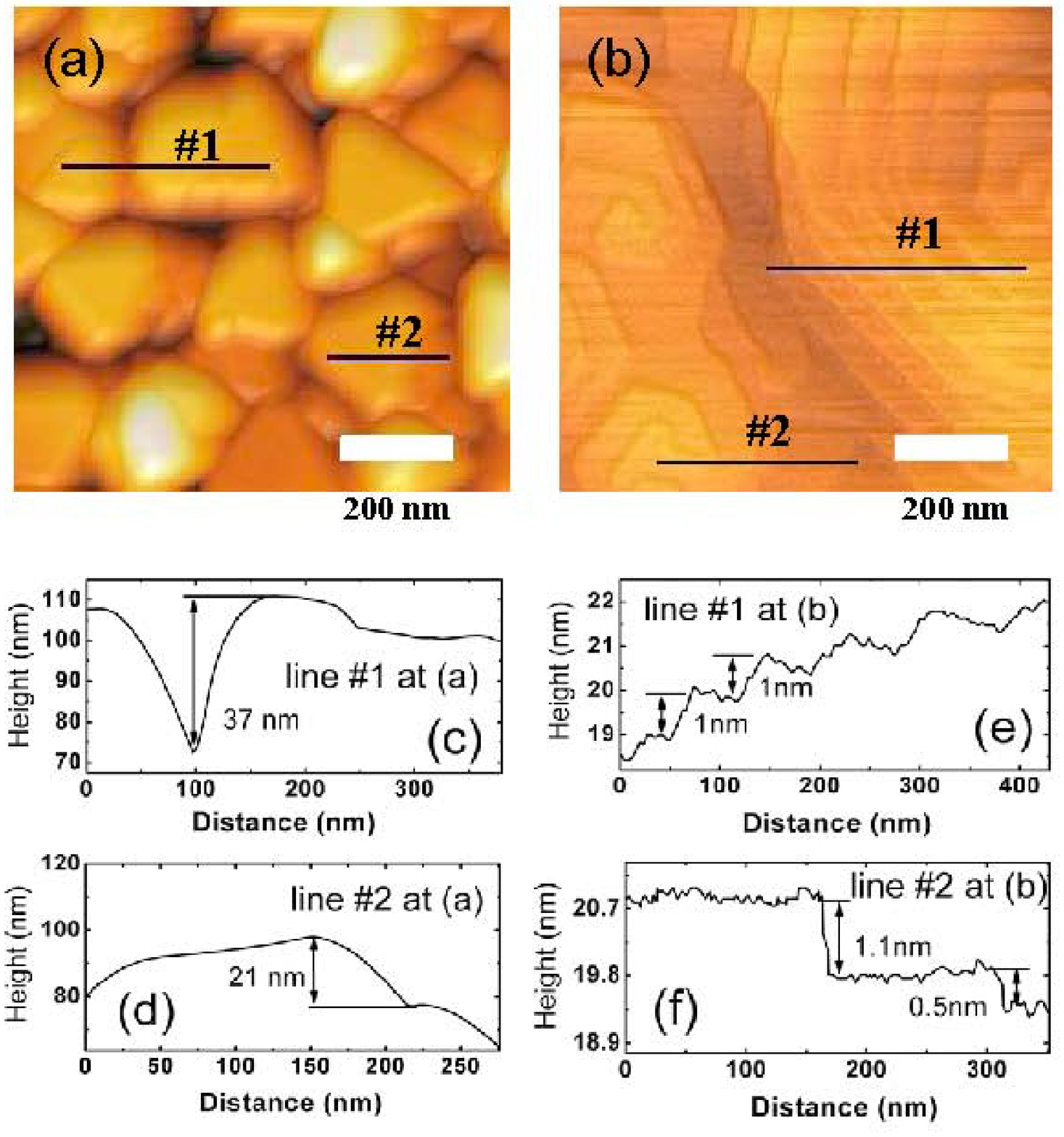}
\caption{ AFM images of (a) $\beta$-Na$_{0.6}$CoO$_{2}$ thin film
and (b) $\gamma$-Na$_{0.7}$CoO$_{2}$ thin film. Sectional contour
graph of (c) the $\#$1-line in $\beta$-Na$_{0.6}$CoO$_{2}$ thin film,
(d) the $\#$2-line in $\beta$-Na$_{0.6}$CoO$_{2}$ thin film, (e) the
$\#$1-line in $\gamma$-Na$_{0.7}$CoO$_{2}$ thin film, (f) the $\#$2-line
in $\gamma$-Na$_{0.7}$CoO$_{2}$ thin film. The
$\gamma$-Na$_{0.7}$CoO$_{2}$ thin film shows spiral surface growth
with multi-terraces. The terrace heights in (e) and (f) are nearly
half or same of one lattice unit.}
%\label{f1}
\end{figure}


\newpage

\begin{references}

\bibitem{son1} I. Terasaki, Y. Sasago, and K. Uchinokura, Phys. Rev. B {\bf 56}, 12685 (1997).


\bibitem{son2} T. Motohashi, E. Naujalis, R. Ueda, K. Isawa, M. Karppinen, and Yamauchi, H.  Appl. Phys. Lett, {\bf 79}, 1480 (2001).


\bibitem{son3} Y. Wang, Nyriss S. Rogado, R. J. Cava, and N. P. Ong, Nature {\bf 423}, 425 (2003).

\bibitem{son4} Kazunori Takada, Hiroya Sakurai, Eiji Takayama-Muromachi, Fujio Izumi, Ruben A. Dilanian, and Takayoshi Sasaki, Nature {\bf 422}, 53 (2003).



\bibitem{son5} F. C. Chou, J. H. Cho, P. A. Lee, E. T. Abel, K. Matan, and Y. S. Lee, Phys. Rev. Lett. {\bf 92}, 157704 (2004).

\bibitem{son6} Q. Huang. B. Khaykoichi, F. C. Chou, J. H. Cho, J. W. Lynn, and Y. S. Lee, Phys. Rev. B {\bf 70}, 134115 (2004)


\bibitem{son7} N. L. Wang, P. Zheng, D. Wu, Y. C. Ma, T. Xiang, R. Y. Jin, and D. Mandrus, Phys. Rev. Lett. {\bf 93}, 237007 (2004).


\bibitem{son8} R. Jin, B. C. Sales, P. Khalifah, and D. Mandrus, Phys. Rev. Lett. {\bf 56}, 217001 (2003).

\bibitem{son9} Maw Lin Foo, Yayu Wang, Satoshi Watauchi, H. W. Zandbergen, Tao He, R. J. Cava, and N. P. Ong, Phys. Rev. Lett. {\bf 92}, 247001 (2004).



\bibitem{son10}  R. J. Balsys Davis. Solid State Ionics {\bf 93}, 279 (1996).

\bibitem{son11} Yasuhiro Ono, Ryuji Ishikawa, Yuzuru Miyazaki, Yoshinobu Ishii, Yukio Morii, and Tsuyoshi Kajitani, J. Solid State Chem, {\bf 166}, 177 (2002).


\bibitem{son12} C. Thinaharan, D. K. Aswal, A. Singh, S. Bhattacharya, N. Joshi, S. K. Gupta, and J. V. Yakhmi, Cryst. Res. Technol. {\bf 39}, 572 (2004).


\bibitem{son13} Alain Karma and Mathis Plapp, Phys. Rev. Lett. {\bf 81}, 4444 (1998).

\end{references}
\end{document}